\documentclass[british]{Interspeech2024_arxiv}

\usepackage[utf8]{inputenc}
\usepackage[T1]{fontenc}
\usepackage[british]{babel}

\usepackage{enumitem} 
\setlist{nolistsep} 

\usepackage{multirow} 
\usepackage{rotating} 

\interspeechcameraready

\title{Should you use a probabilistic duration model in TTS? Probably!\\{}Especially for spontaneous speech}

\name[affiliation={1}]{Shivam}{Mehta}
\name[affiliation={1}]{Harm}{Lameris}
\name[affiliation={2}]{Rajiv}{Punmiya}
\name[affiliation={1}]{Jonas}{Beskow}
\name[affiliation={1}]{\'Eva}{Sz\'ekely}
\name[affiliation={1}]{Gustav Eje}{Henter}

\address{
  $^1$Speech, Music and Hearing, KTH Royal Institute of Technology, Sweden \hfill $^2$Independent researcher}
\email{\href{mailto:smehta@kth.se}{smehta@kth.se}, \href{mailto:lameris@kth.se}{lameris@kth.se}, \href{mailto:rajiv.punmiya@gmail.com}{rajiv.punmiya@gmail.com}, \href{mailto:beskow@kth.se}{beskow@kth.se}, \href{mailto:szekely@kth.se}{szekely@kth.se}, \href{mailto:ghe@kth.se}{ghe@kth.se}}


\usepackage{pdfrender}
\newcommand\tablebf[1]{\textpdfrender{TextRenderingMode=FillStroke,LineWidth=0.5}{#1}} 

\newcommand{\detwins}[1]{\textit{#1}}
\newcommand{\fmwins}[1]{\tablebf{#1}}

\usepackage{graphicx}
\usepackage{xcolor}
\graphicspath{{./figures/}}
\DeclareGraphicsExtensions{.pdf,.png,.jpeg,.jpg}

\begin{document}

\maketitle

\begin{abstract}
Converting input symbols to output audio in TTS requires modelling the durations of speech sounds. Leading non-autoregressive (NAR) TTS models treat duration modelling as a regression problem. The same utterance is then spoken with identical timings every time, unlike when a human speaks. Probabilistic models of duration have been proposed, but there is mixed evidence of their benefits. However, prior studies generally only consider speech read aloud, and ignore spontaneous speech, despite the latter being both a more common and a more variable mode of speaking. We compare the effect of conventional deterministic duration modelling to durations sampled from a powerful probabilistic model based on conditional flow matching (OT-CFM), in three different NAR TTS approaches: regression-based, deep generative, and end-to-end. Across four different corpora, stochastic duration modelling improves probabilistic NAR TTS approaches, especially for spontaneous speech.
\end{abstract}

\section{Introduction}

A key challenge of text-to-speech is upsampling discrete text inputs, usually graphemes or phonemes, into continuous-valued acoustic outputs, often in the form of mel-spectrograms.
It is of great importance to accurately model speech-sound durations in this upsampling process, particularly for the prosody of the speech.
Traditionally, autoregressive (AR) neural TTS models
infer these durations implicitly within their generative process, whilst non-autoregressive (NAR) TTS models
often require an explicit model to create these durations.
Such duration models can adopt either a deterministic regression-based approach, producing the same output for constant input, or a stochastic framework, learning a probability distribution and generating different samples from that distribution.
Despite a foundation of theoretical \cite{gretton2012kernel} and experimental evidence \cite{henter2014} highlighting
that only probabilistic synthesis methods can appear perfectly natural,
it is only recently that advances in probabilistic modelling have demonstrated standout results in synthesising
human behaviour such as motion \cite{alexanderson2023listen, mehta2023diff, mehta2024unified} and speech \cite{mehta2024matcha, popov2021grad, chen2021wavegrad2, kim2021vits, chen2021wavegrad, kongdiffwave, popov2021diffusion}.

Despite the apparent advantages of stochastic approaches in various domains of synthetic content generation, the adoption of probabilistic duration modelling in NAR TTS remains limited. At present, a majority of widely used NAR TTS models employ regression-based, deterministic duration modelling. This includes not only regression-based approaches like \cite{lancucki2021fastpitch, ren2021fastspeech2} but also prominent examples of the latest technological advancements like diffusion models \cite{popov2021grad} and flow-matching-based models \cite{le2023voicebox, guo2024voiceflow, mehta2024matcha}. The end-to-end TTS model VITS \cite{kim2021vits, kong2023vits2} is an exception that uses stochastic duration modelling.

This reluctance to employ stochastic duration modelling can be partly attributed to mixed empirical evidence regarding its efficacy. Some studies \cite{kim2021vits, kong2023vits2} find stochastic duration modelling to improve 
the naturalness of synthesised speech, whilst more recent ones challenge its effectiveness \cite{le2023voicebox}. Moreover, evaluations comparing deterministic and stochastic approaches often limit their focus to read-aloud speech corpora like LJ Speech \cite{ljspeech17}.
This leaves open the question of how durations are to be modelled in spontaneous speech, in light of its highly diverse prosodic structure \cite{lameris2023prosody} and that it constitutes the most common form of human speech communication.

In this paper, we perform a comprehensive comparison
between conventional regression-based duration modelling and durations sampled from a powerful probabilistic duration model based on flow matching. We perform this comparison across a variety of NAR TTS architectures, specifically a deterministic acoustic model (FastSpeech 2 \cite{ren2021fastspeech2}), an advanced deep generative acoustic model (Matcha-TTS \cite{mehta2024matcha}), and a probabilistic end-to-end TTS model (VITS \cite{kim2021vits}). For each architecture, deterministic and stochastic duration modelling is evaluated both objectively and through subjective listening tests on a total of four different speech corpora: two comprising read-aloud speech and two containing spontaneous speech.
Our key findings are:
\begin{itemize}
    \item Regression-based TTS approaches do not benefit from stochastic duration modelling. In contrast, the probabilistic TTS approaches had equal or improved performance.
    \item The differences between deterministic and probabilistic duration modelling are most evident in spontaneous speech corpora, and least apparent in the widely used LJ Speech corpus. This highlights the need for better benchmarks of how modern TTS systems handle the complexities of natural speech.
\end{itemize}
Our findings indicate that stochastic modelling of speech-sound durations can improve NAR TTS, and that flow-matching models introduce negligible overhead in terms of parameter count and synthesis speed in this regard.
For audio and resources see \url{https://shivammehta25.github.io/prob_dur/}.

\section{Background}

\subsection{Duration modelling in TTS}


Contemporary TTS models are generally classified as either autoregressive or non-autoregressive.
Autoregressive models generate output sequentially, using either neural attention mechanisms (e.g., \cite{shen2018natural, li2019neural_transformer_tts}) or transducers (e.g., \cite{mehta2022neuralhmm, mehta2023overflow}) to upsample input symbols or states to output frames as they go along.
Non-autoregressive models, in contrast, generate all output values in parallel.
This can be faster, especially on GPUs. 
These models typically upsample the input text vectors using an explicit duration model, after which the vectors are transformed into acoustic features.
The duration models are trained on reference durations obtained either through external forced alignment \cite{ren2021fastspeech2, lancucki2021fastpitch}, or via Monotonic Alignment Search (MAS) \cite{kim2020glow, popov2021grad, mehta2024matcha, guo2024voiceflow, kim2023pflow}, or determined through Gaussian upsampling \cite{shen2020non, chen2021wavegrad2}.
All these systems employ a regression-based duration predictor trained to typically minimise the log-domain mean square error (MSE) between predicted and reference durations.
All the cited works primarily perform experiments on read-aloud speech.

Notably, there are existing studies on read speech \cite{ogun2023stochastic} demonstrating that stochastic prosody modelling contributes to addressing the issue of oversmoothness in synthesis, particularly for styles characterised by high F0 \cite{zhang2023comparing}. However, these investigations did not isolate the influence of the duration modelling from other conditioning inputs like F0, nor did they examine its effects on highly conversational, spontaneous speech. Moreover, they often rely on generative modelling paradigms with restricted expressiveness, that require numerous iterations or neural network calls to produce good quality samples, which introduces significant computational overhead.
Specific examples of this are VITS \cite{kim2021vits}, which uses discrete-time normalising flows conditioned on speaker embeddings to generate durations, and VITS 2 \cite{kong2023vits2}, which wraps the aforementioned duration predictor with a discriminator to produce more realistic duration values.
However, they only train the duration model after the acoustic model, as opposed to
jointly with the rest of the model, for reasons of training stability.

Recent studies \cite{le2023voicebox} have hinted that regression-based duration predictors underestimate the standard deviation of phoneme and silence durations, producing samples with less duration diversity and more regular durations.
This could potentially be desirable for synthesising read-aloud speech, but for realistic conversational synthesis we need to recreate the variability and irregularity in phoneme and silence durations.


Recently, a new class of probabilistic generative modelling known as conditional flow matching \cite{lipman2023flow} (specifically OT-CFM and the closely related rectified flows \cite{liu2022flow}) have emerged in the text-to-speech domain, delivering fast and state-of-the-art results in acoustic modelling \cite{le2023voicebox, mehta2024matcha, guo2024voiceflow, kim2023pflow}. Flow matching is a continuous-time variant of normalising flows that, unlike discrete-time normalising flows \cite{kingma2018glow}, can be trained without ODE solvers and does not require limiting the architectural design to ensure bijectivity, improving training speed and model flexibility.
The specific design of OT-CFM means that very few network evaluations are needed at synthesis time, making it much faster than diffusion models \cite{lipman2023flow}.
Despite this, only \cite{le2023voicebox} among the cited works investigates flow matching for TTS duration modelling, finding no notable naturalness improvement.

\subsection{Spontaneous speech}
Spontaneous speech is the most common form of speech humans produce and comprehend \cite{shriberg2005spontaneous}, yet it has considerably different characteristics to the read-aloud speech that state-of-the-art TTS models are generally trained on. It offers interesting challenges for TTS, as it has more diversity in its F0 and speech rate compared to read speech, even if the read speech has a conversational style \cite{lameris2023prosody}. Additionally, it leverages phenomena not found in scripted speech, such as disfluencies, filled pauses, and breaths, all essential tools in speech planning and beneficial to information recall \cite{szekely2020breathing}. There is also considerably more variability in spontaneous speech depending on the communicative context, with prosodic realisation playing a role in the interpretation of the pragmatic implication of a phrase \cite{herment2012pragmatic}.

With the increased focus on communicative competencies in spoken interaction for agents, spontaneous speech offers a more realistic setting for the evaluation of TTS architectures \cite{marge2022spoken}. However, despite its potential advantages, spontaneous speech data has not been commonly used in TTS systems, partly due to its expensive transcription process. The lower variability of read-aloud speech also benefited concatenative TTS systems by making differences across concatenation points smaller.
As a consequence, there is little knowledge regarding what techniques that are appropriate for synthesising spontaneous speech.
In particular, we believe our work is the first to carefully study duration modelling in spontaneous speech for NAR neural TTS.

\section{Method}
\subsection{Data}
\label{ssec:data}
We used two read-speech and two spontaneous-speech corpora for our study of duration modelling. 100 utterances of each corpus were withheld for validation.

The first read-speech corpus, LJ Speech (herein labelled \textbf{LJ}) \cite{ljspeech17}, is a public-domain dataset that consists of a single female speaker of General American English reading passages from non-fiction books for a duration of approximately 24~h. The second read-speech corpus, RyanSpeech \textbf{(RS)} \cite{zandie2021ryanspeech}, is a scripted conversational corpus of 9~h of a male speaker of General American English reading texts from chatbots and dialogue systems, as well as LibriVox transcriptions. 
Whilst purportedly conversational,
RyanSpeech does not exhibit any spontaneous speech behaviours, as the performed dialogues are all scripted. 

The experiments considered two spontaneous-speech corpora. One was the Trinity
Speech-Gesture Dataset II (\textbf{TSGD2})\footnote{\url{https://trinityspeechgesture.scss.tcd.ie/}} \cite{ferstl2021expressgesture}, a 6~h spontaneous conversational corpus of time-aligned speech and marker-based motion capture of a male Hiberno-English speaking actor. It features 25 takes of the actor speaking in a conversational style without feedback or interruptions.
The speech data was extracted and
segmented into breath groups of 1--10 seconds, which were transcribed using Automatic Speech Recognition (ASR) before manual correction, based on \cite{szekely2019casting}.
The transcripts include fillers and repetitions, along with tokens like semicolons for breaths and commas for silent pauses, to be able to elicit spontaneous behaviours at synthesis time.

Our other spontaneous corpus, AptSpeech (\textbf{AptS}), contains the speech data from the publicly available multimodal multi-party recordings from \cite{kontogiorgos2018multimodal}. The corpus consists of 15 interactions between a single mediator and varying participants who were tasked with designing a living space (apartment) on a large touchscreen GUI. We extracted the unscripted speech data from the mediator, a male speaker of General American English, which totalled 5.7~h.
Transcription and segmentation were performed identically to the first spontaneous corpus.

\subsection{Duration models considered}
\label{ssec:duration_models}
\begin{figure}[!t]
    \centering
    \includegraphics[width=\columnwidth]{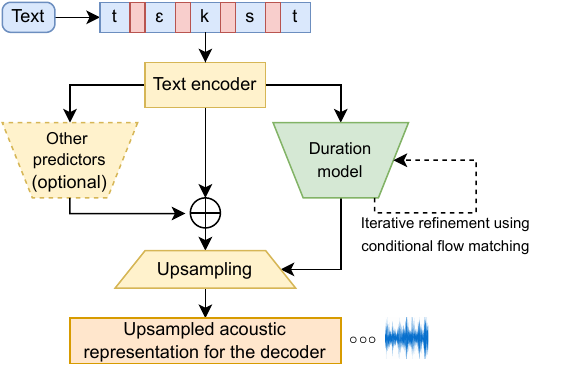}
    \caption{Overview of NAR TTS synthesis. We replace duration prediction with flow-matching-based duration modelling.}
    \label{fig: general architecture}
    \vspace{-1.5\baselineskip}
\end{figure}

Most contemporary NAR TTS approaches are encoder-decoder frameworks where duration modelling occupies a consistent position in the synthesis pipeline, illustrated in Figure \ref{fig: general architecture}.
Specifically, the encoder generates an intermediate representation from the input symbols, which is fed into a duration-predictor module. This module estimates the temporal length (i.e., the number of frames) of each unit in the input sequence. 
%
%
The duration model is trained to predict the reference duration of each input symbol using a MSE regression loss in the log domain \cite{ren2019fastspeech}.
It thus estimates 
the expected log-duration of each unit, conditioned on the text-encoder output sequence.
After converting durations to integers
each text-encoder output vector is repeated (upsampled) as many times as the duration predictor indicates.

We explore the effects of replacing the MSE-based duration predictor in existing NAR TTS approaches with a log-domain duration model based on conditional flow matching, specifically OT-CFM \cite{lipman2023flow} with $\sigma{=}10^{-4}$ as in \cite{mehta2024matcha}.
This learns a probability distribution by learning to predict log-domain reference durations (no dequantisation) from noisy versions of the same, and from the text-encoder output.
Synthesis iteratively transforms a sequence of Gaussian noise values into output durations, then converts to the linear domain and rounds to the nearest integer. 

\section{Experiments}
For our experiments, we selected three strong and widely-recognised NAR TTS approaches: FastSpeech 2 (\textbf{FS2}), a deterministic acoustic model; Matcha-TTS (\textbf{Matcha}), a probabilistic acoustic model; and VITS (\textbf{VITS}), an end-to-end probabilistic TTS model.
For each approach, we studied the effect of replacing its conventional deterministic duration predictor (\textbf{DET}) with the OT-CFM-based duration model (\textbf{FM}) described in Sec.\ \ref{ssec:duration_models}.
Whilst Matcha-TTS and VITS can learn alignments jointly with learning to speak, FastSpeech 2 was supplied with pre-computed reference alignments from Matcha during its training phase. We trained each of these architectures with each duration-model type (DET and FM) on the four different corpora from Sec.\ \ref{ssec:data}.
for 500k updates. This resulted in a total of 3$\times$2$\times$4=24 distinct systems trained. Each system was trained on a single NVIDIA-3090 GPU using batch size 32.

All systems used Phonemizer\footnote{\url{https://github.com/bootphon/phonemizer/}} with \texttt{espeak-ng} to convert input graphemes to IPA phones. Following \cite{kim2020glow, kim2021vits, mehta2024matcha}, we consistently interleaved each phone with a blank token, so as to represent each phone by two encoder vectors.
This improves acoustic-model granularity.
We used the default hyperparameters for each architecture as specified in their original publications, with the sole deviation being the incorporation of an additional convolution-based causal post-net for FS2, a feature prevalent in most open-source implementations of FastSpeech 2.\footnote{\url{https://github.com/facebookresearch/fairseq/}} Notably, all approaches
use the same network architecture for the duration predictor, consisting of a layer norm sandwiched between two convolution layers followed by a projection layer, cumulatively amounting to approximately 400k parameters.

\begin{figure}[!t]
    \centering
    \includegraphics[width=\columnwidth]{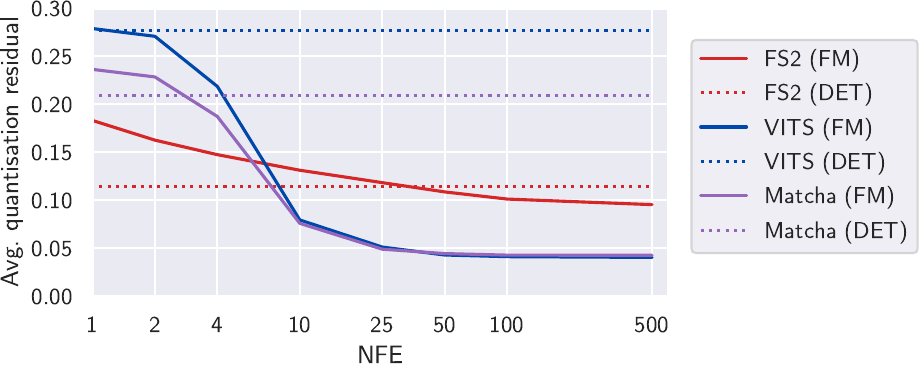}
    \caption{Duration quantisation residual, averaged over four corpora for FS2, VITS, and Matcha-TTS vs.\ the number of function evaluations for FM. DET has constant residual.}
    \label{fig:quantisation residual}
    \vspace{-1.5\baselineskip}
\end{figure}

For the FM duration models, we maintained the architecture of the original, deterministic duration predictor unchanged but
incorporated noisy durations and an embedding of the current iteration step as extra inputs,
the latter using the time-step embedding architecture from the acoustic decoder of \cite{mehta2024matcha}.
This only added another 100k parameters, equating to 0.6\% or less of model size. We empirically tuned the noise temperature to 0.667 during synthesis, finding that VITS (but not Matcha) output degraded above this value.
We opted to use 10 Neural Function Evaluations (NFEs) during synthesis, which resulted in a negligible increase in of 0.001 in Real Time Factor (RTF) over DET.
Figure \ref{fig:quantisation residual} graphs the average magnitude of the residual when quantising the duration-model output to the nearest integer as a function of on the NFE, indicating how accurately the synthesis process is able to generate integer outputs.

\subsection{Stimuli and objective evaluation}
\label{sec:stim}
To ensure that all stimuli are understandable as standalone utterances and their content is appropriate for online listening tests -- which is not true for most corpora -- we generated 100 new test sentences per dataset in the same style as the original corpus (to avoid domain mismatch).
This was done by putting 100 sentences from each corpus into GPT-4 \cite{achiam2023gpt}, with instructions to mimic the speaking style and breaths and pauses for the spontaneous corpora, and create another 100 sentences related to everyday topics like visiting the zoo, going to a shopping centre, and going to school.
All prompts and sentences are provided on our webpage.
Five random realisations per sentence were synthesised for the stochastic duration models, with objective scores being calculated as an average across all of these, whilst the subjective evaluation used one realisation per sentence.

For the objective evaluation, we calculated the word error rate (WER) of automatic speech recognition on synthetic stimuli, and also performed automatic MOS prediction to estimate TTS quality.
WERs were obtained using Whisper \texttt{medium.en} \cite{whisper}, as the WER of contemporary ASR correlates well with speech intelligibility to human listeners \cite{taylor2021confidence}.
MOS prediction used the off-the-shelf AutoMOS system described in \cite{cooper2022generalization}.

\subsection{Subjective evaluation}
For the subjective evaluation, we conducted four Comparative Mean Opinion Score (CMOS)-style web-based listening tests, with each test including stimuli from all models, but only one specific corpus. In these tests, listeners were asked to ``choose how natural these two versions sound in comparison''
on a 7-point integer Likert scale ranging from ``ver 1 much better'' to ``ver 2 much better''; zero meant no difference.
Each audio stimulus pair compared the same sentence synthesised using the same architecture (FS2, VITS, or Matcha), the only difference being the use of either DET or FM for durations.
We recruited 160 self-reported native English speakers via Prolific
, exactly 40 for each test.
Each listener was paid 3 GBP after test completion (median duration 15 min).
An additional 7 listeners were rejected for failing two or more of our three attention checks.

To familiarise listeners with the speaker and speaking style of the corpus used in the test, they first listened five natural utterances from the test set, before proceeding with the CMOS test. Each CMOS test for a specific listener comprised 10 stimulus pairs per architecture (30 pairs per person).
Ultimately 4 sets of 30 stimulus pairs per corpus were evaluated, for an overall total of 400 ratings for each of the 12 DET-FM system pairings.

\section{Results}

\begin{table}[!t]
\centering
\caption{ASR WER, AutoMOS scores,
and CMOS values. Positive CMOS means FM is preferred over DET. Significant CMOS differences favouring FM are bold, ones favouring DET italic.}
\label{tab:results}
\begin{tabular}{@{}l|l|cc|cc|c@{}}
\toprule
Model & \multicolumn{1}{c|}{Dataset} & \multicolumn{2}{c|}{WER\% ($\downarrow$)}                                     & \multicolumn{2}{c|}{AutoMOS ($\uparrow$)}                                 & \multicolumn{1}{c}{CMOS}        \\
      &                              & \multicolumn{1}{c}{DET} & \multicolumn{1}{c|}{FM}            & \multicolumn{1}{c}{DET} & \multicolumn{1}{c|}{FM}            & 95\% conf.\ int.                                \\ \midrule
\multirow{4}{*}{\begin{turn}{90}
FS2
\end{turn}}
& LJ                           & \hphantom{0}\tablebf{2.07}           & \hphantom{0}2.27                               & \tablebf{4.35}           & 4.29                               & \detwins{$-$0.25$\pm$0.15}\\
      & RS                           & \hphantom{0}2.89                    & \hphantom{0}\tablebf{2.27}                      & \tablebf{4.35}           & 4.29                               & \detwins{$-$0.49$\pm$0.13}\\
      & TSGD2                        & \tablebf{17.24}          & 23.52                              & \tablebf{2.65}           & 2.44                               & \detwins{$-$0.59$\pm$0.10}\\
      & AptS                         & \tablebf{27.61}          & 34.17                              & \tablebf{3.15}           & 2.88                               & \detwins{$-$1.66$\pm$0.11}\\ \midrule
\multirow{4}{*}{\begin{turn}{90}
VITS
\end{turn}}
& LJ                           & \hphantom{0}2.67                    & \hphantom{0}\tablebf{2.40}                      & 4.31                    & \tablebf{4.51}                      & \hphantom{$-$}\fmwins{0.23$\pm$0.17} \\
      & RS                           & \hphantom{0}2.46                    & \hphantom{0}\tablebf{2.31}                      & 4.52                    & \tablebf{4.70}                      & \hphantom{$-$}\fmwins{0.47$\pm$0.16} \\
      & TSGD2                        & 13.27                   & \hphantom{0}\tablebf{9.26}                      & 3.38                    & \tablebf{3.94}                      & \hphantom{$-$}\fmwins{0.48$\pm$0.15}\\
      & AptS                         & 10.62                   & \hphantom{0}\tablebf{8.31}                      & 4.07                    & \tablebf{4.37}                      & \hphantom{$-$}\fmwins{0.69$\pm$0.14}\\ \midrule
\multirow{4}{*}{\begin{turn}{90}
Matcha
\end{turn}}
& LJ                           & \hphantom{0}2.39                    & \multicolumn{1}{c|}{\hphantom{0}\tablebf{1.53}} & 4.40                     & \multicolumn{1}{c|}{\tablebf{4.53}} & \hphantom{$-$}0.02$\pm$0.14    \\
      & RS                           & \hphantom{0}1.94                    & \multicolumn{1}{c|}{\hphantom{0}\tablebf{1.70}} & 4.46                    & \multicolumn{1}{c|}{\tablebf{4.56}} & $-$0.04$\pm$0.16   \\
      & TSGD2                        & \hphantom{0}9.00                       & \multicolumn{1}{c|}{\hphantom{0}\tablebf{6.15}} & 3.12                    & \multicolumn{1}{c|}{\tablebf{3.56}} & \hphantom{$-$}\fmwins{0.47$\pm$0.15} \\
      & AptS                         & 11.50                    & \multicolumn{1}{c|}{\hphantom{0}\tablebf{5.11}} & 3.55                    & \multicolumn{1}{c|}{\tablebf{4.15}} & \hphantom{$-$}\fmwins{0.69$\pm$0.14} \\ \bottomrule
\end{tabular}
\vspace{-\baselineskip}
\end{table}
Results of the objective and subjective experiments are reported in Table \ref{tab:results}.
All CMOS results except LJ and RS for Matcha were significantly different from zero ($p$ < 0.05).
We find that FS2, the regression-based TTS model, produced worse speech in all aspects when changing from the deterministic to the stochastic duration model.
Conversely, VITS (an end-to-end  architecture based on discrete-time normalising flows) demonstrated notable improvements with the stochastic duration predictor. This enhancement was observed for both read and spontaneous speech, aligning with the findings in the original VITS paper \cite{kim2021vits}. Matcha-TTS, being architecturally flexible with a strong probabilistic foundation, synthesised read speech similarly well using both DET and FM (consistent with prior results on read speech in \cite{le2023voicebox}), but showed significant improvement for spontaneous speech with stochastic duration generation.

It is interesting to compare the numerical results to the quantisation residuals in Figure \ref{fig:quantisation residual}.
These residuals indicate that the stochastic FM models successfully learnt to produce near-integer outputs during sampling, but that 10 or more NFEs (i.e., Euler-forward ODE-solver steps) are required to recover this property during sampling; this validates our choice of NFE=10 for the experiments.
The DET models only require a single NFE to compute their output.
As they are trained to predict average values, which typically are not integers, it makes sense that they have greater quantisation residuals (dashed lines).
The one exception is FS2, where FM duration modelling converges much more slowly and to a higher residual value.
Although the graph only shows averages, the same shapes and trends hold for all individual corpora.
This suggests that OT-CFM struggled to learn accurate duration distributions inside the FS2 architecture, making it perform worse than DET there.
Separately,
the WER and AutoMOS results suggest that FS2 typically achieved worse intelligibility and speech quality than corresponding VITS and Matcha systems.
FS2 FM might improve by using a synthesis temperature near zero, but this essentially makes output deterministic like DET, defeating the purpose of using flow matching.


Our results indicate that, although deterministic acoustic models may not benefit from stochastic duration modelling, probabilistic TTS approaches often improved and were never adversely affected (some cases showed no statistically significant difference). This reinforces the potential benefits of advanced probabilistic duration modelling, especially using low-overhead flow-matching techniques. Further, the disparity between deterministic and probabilistic modelling paradigms was most pronounced on spontaneous speech corpora, underscoring the critical role of accurate duration modelling (and potentially broader prosody modelling) for this diverse and irregular speech type.
This finding, combined with the observed advantages in models like VITS and Matcha-TTS, suggest that spontaneous speech and probabilistic models are promising direction for future TTS research focused on enhancing the naturalness and expressiveness of synthesised speech.

Interestingly, LJ Speech, the most widely used TTS corpus, consistently exhibited the smallest CMOS difference, highlighting its limitations as a benchmark for duration modelling, and possibly other prosodic features as well.
This strongly suggests a need for more varied and rigorous benchmarks and corpora to accurately evaluate the performance of advanced TTS methods.



\section{Conclusion}

In this study, we explored the impact of deterministic versus probabilistic duration modelling within the framework of non-autoregressive TTS, across three categories of such TTS architectures: regression-based, deep generative, and end-to-end.
Crucially, our study included not only read-aloud but also spontaneous speech, for which duration modelling has hitherto been severely underexplored.
We discovered significant naturalness improvements from probabilistic duration modelling over conventional deterministic duration prediction 
on spontaneous speech synthesised by probabilistic TTS paradigms.
Objective metrics likewise improved.
Our collective findings highlight the 
advantages of probabilistic models of durations and acoustics in achieving lifelike and expressive speech synthesis.

Additionally, our analysis suggests that the LJ Speech corpus may not be the most appropriate benchmark for evaluating the nuances of duration and prosody modelling, due to its limited representation of the variability and complexities inherent in natural speech.
Spontaneous speech corpora may offer more interesting and relevant material for benchmarking TTS.

\section{Acknowledgements}
Research funded by the Wallenberg AI, Autonomous Systems and Software Program (WASP) funded by the Knut and Alice Wallenberg Foundation, Swedish Research Council proj.\ VR-2020-02396, and the Industrial Strategic Technology Development Program (grant no.\ 20023495) funded by MOTIE, Korea.
\bibliographystyle{IEEEtran}
\bibliography{mybib}

\end{document}